# Quantifying the Computational Capability of a Nanomagnetic Reservoir Computing Platform with Emergent Magnetisation Dynamics

I T Vidamour[1], M O A Ellis[2], D Griffin[3], G Venkat[1], C Swindells[1], R W S Dawidek[1], T J Broomhall[1], NJ Steinke[4], J F K Cooper[4], F Maccherozzi[5], S S Dhesi[5], S Stepney[3], E Vasilaki[2,6], D A Allwood[1], T J Hayward[1]

1- Department of Materials Science and Engineering, University of Sheffield, Sheffield S1 3JD, United Kingdom
2- Department of Computer Science, University of Sheffield, Sheffield S1 4DP, United Kingdom
3- Department of Computer Science, University of York, York YO10 5GH, United Kingdom
4- ISIS Neutron and Muon Source, Rutherford Appleton Lab, Didcot, OX11 0QX, United Kingdom
5- Diamond Light Source, Harwell Science and Innovation Campus, Didcot, Oxfordshire OX11 0DE, United Kingdom
6- Institute of Neuroinformatics, University of Zurich and ETH Zurich, 8057 Zürich, Switzerland

**Abstract**

Devices based on arrays of interconnected magnetic nano-rings with emergent magnetization dynamics have recently been proposed for use in reservoir computing applications, but for them to be computationally useful it must be possible to optimise their dynamical responses. Here, we use a phenomenological model to demonstrate that such reservoirs can be optimised for classification tasks by tuning hyperparameters that control the scaling and input-rate of data into the system using rotating magnetic fields. We use task-independent metrics to assess the rings' computational capabilities at each set of these hyperparameters and show how these metrics correlate directly to performance in spoken and written digit recognition tasks. We then show that these metrics, and performance in tasks, can be further improved by expanding the reservoir's output to include multiple, concurrent measures of the ring arrays' magnetic states.

**Introduction**

Neuromorphic devices use inherent material properties to perform brain-like computational operations *in materio*. This allows for improvements in efficiency over standard artificial neural networks as neural architectures are directly *emulated* in hardware, rather than *simulated* using conventional computers [1].

Reservoir computing (RC) is a machine learning paradigm that is well-suited to *in materio* implementations. In RC, a fixed dynamical system (the reservoir) transforms input signals into higher dimensional representations, facilitating classification in cases where input data is linearly inseparable. In the archetypal Echo State Networks (ESNs) [2], the reservoir takes the form of a recurrent neural network (RNN) initialised with a sparse, random connectivity matrix. A linear readout layer provides output from the weighted sum of activity across nodes within the reservoir [3]. ESNs address the well-known difficulties of training RNNs, and recent models have improved both their applicability to classification tasks and their robustness against catastrophic forgetting [4].

While ESNs are typically simulated on conventional computers, recent studies have shown that computational ability is preserved if the RNNs are replaced by physical systems [5]–[12] with the correct properties: nonlinearity between input and output, and 'fading' memory of past inputs. With typical hardware implementations of reservoir computing there is no separation between the components used for computation and those used for memory, mitigating the von-Neumann bottleneck associated with discrete memory and computation units. This offers potential benefits in terms of reduced latency, increased computational power per unit area, and improved energy efficiency of the system.

Different classes of physical systems have been proposed for RC, each with their own advantages and technological challenges [13]. For example, optoelectronic systems [6], [14] utilise the nonlinear properties of off-the-shelf intensity modulation components coupled with fibre optic spools as delay lines to create reservoirs that can perform computational tasks with performance that rivals the state-of-the-art [14]. However, these suffer from substantial challenges to minimisation due to the size of optical delay lines. Molecular platforms use molecules such as proteins [15] and enzymes[16] as the computational building blocks for RC. These offer advantages in terms of the complexity of the reservoir and the feasibility of miniaturisation, though throughput is often slow and interfacing with standard electrical components is challenging. Memristors have also been proposed for RC [17], [18], as well as other forms of computation [19]. Memristors use short-term memory effects created from their variable resistance over time and are a particularly promising implementation of RC due to their suitability for interfacing with standard electronics as well as the ease of miniaturisation.

Nanomagnetic platforms are also well-suited to creating hardware-based reservoirs, offering many desirable properties including non-volatility, which provides a natural path to memory, and inherent non-linearity in their dynamics. Furthermore, methods for electrical reading [20], [21] and writing [22], [23] data are well-established from both the development of commercial magnetic random access memory (MRAM) [24], as well as research into more novel nanomagnetic logic [25]–[27] and memory [28], [29] devices. A wide range of nanomagnetic systems have been proposed for use as reservoirs including spin-torque oscillators (STOs) [12], super-paramagnetic arrays [10], skyrmion textures [30], single domain walls [11], artificial spin-ices [8], and garnet films [9].

Recently, we have proposed arrays of interconnected magnetic nano-rings as candidate platforms for nanomagnetic RC (Figure 1(a)) [31]. The arrays, which are lithographically patterned from thin films, consist of planar, ring-shaped nanowires of the soft magnetic material $Ni_{80}Fe_{20}$ with typical ring diameters <5 µm, linewidths <500 nm and film thicknesses <40 nm. For these dimensions the magnetic ground state of the rings in the array are "vortex" states (Figure 1(a)), where the local magnetization vector rotates in a closed loop around the rings' circumferences. However, they can also support meta-stable, bi-domain "onion" states where magnetic domains with anti-parallel circulation are separated by a pairs of magnetic domain walls (DWs) (Figure 1(b)).

The soft magnetic properties of the nanowires means that the DWs are highly mobile and propagate through the nanowires like rigid quasi-particles when subjected to applied magnetic fields [17]. Specifically, in-plane rotating fields can drive the DWs pairs to coherently and continuously rotate around the ring circumferences [32] (Figure 1(c)). While in isolated rings DW motion is relatively unimpeded, in interconnected ring arrays junctions between the rings act as pinning sites that present localised energy barriers against DW propagation. The interaction of the DWs with such pinning sites are highly stochastic [28], [29] such that when DWs encounter junctions during their rotation around the ring they have a finite probability of becoming pinned temporarily in place, with pinning becoming less likely as the rotating field amplitude is increased. These pinning events lead to field-dependent stochastic interactions between pinned and propagating DWs at the rings' junctions which can cause both loss of DW pairs from the array (i.e., increasing in the number of vortex states) or gain of DW pairs (i.e., decreasing in the number of vortex states).

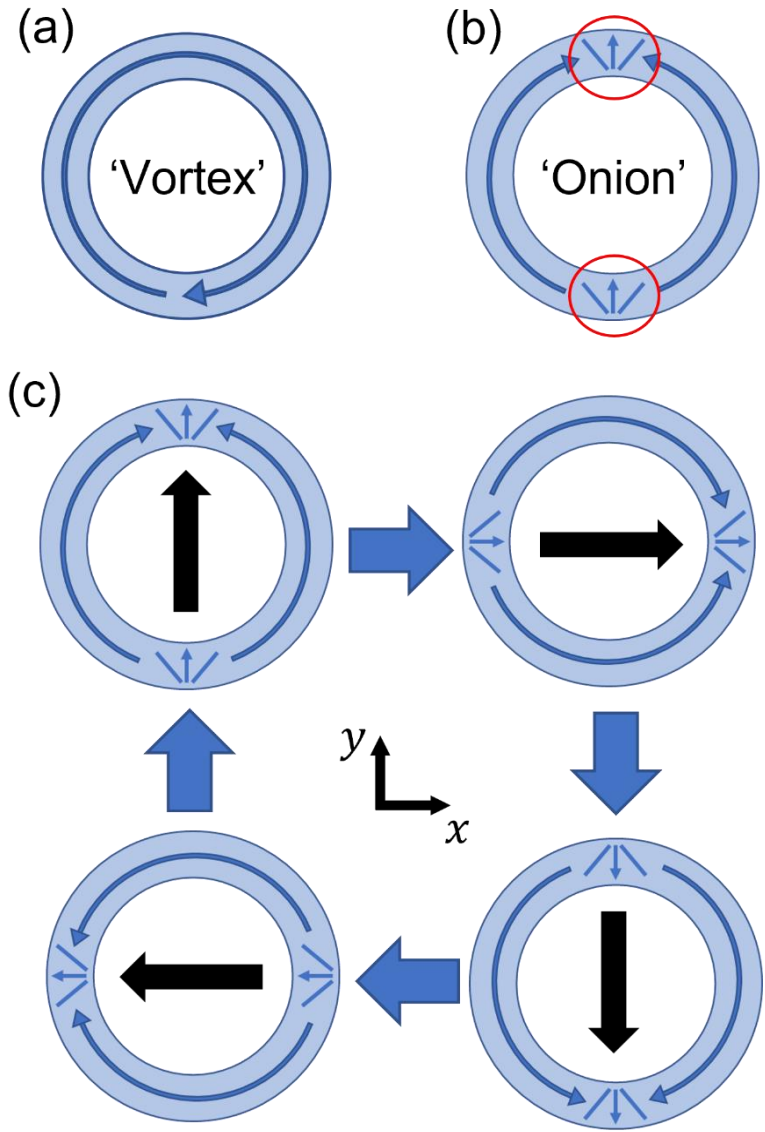

*Figure 1- (a) Schematic diagram of a 'vortex' state. The blue arrows represent the local magnetisation direction (b) An 'onion' state, featuring two domains (separated by a head-to-head domain wall (upper red circle), and a tail-to-tail domain wall (lower red circle). (c) Coherent rotation of domain walls as the applied magnetic field direction (black arrows) rotates through 360°.*

Collectively, these effects result in the ring arrays' magnetisation states exhibiting emergent responses to rotating magnetic fields, where the simple interactions of DWs at the junctions between rings results in complex collective behaviour of the arrays as a whole [31]. In our previous work we have shown that these emergent dynamics result in both a highly non-linear response of an arrays' magnetisation states to the rotating field amplitude, and fading memory of previous magnetisation states, thus meeting the two primary criteria for a dynamical system to be used for RC [31]. The transformations provided by the rings' response can be varied by controlling how input data scales the applied field and its input rate. This offers the possibility of tuning their responses for different computational tasks. Furthermore, it is well established that the magnetisation states of magnetic nanorings can be characterised electrically using either anisotropic magnetoresistance measurements [21], [35], or by giant magnetoresistance if the rings are patterned from multilayer films with spin valve properties [36], [37], making them highly suitable for device integration. Together, these properties suggest interconnected rings have great potential for in-materio reservoir computing.

In a previous work we used a phenomenological model of a ring array's dynamics to demonstrate that these systems could be used as reservoirs [31]. This was achieved by treating an array as a single dynamical node into which time-multiplexed data [5] was input via the amplitude of a rotating applied magnetic field. Our simulations demonstrated that the ring arrays could successfully perform benchmark classification tasks such as spoken digit recognition. However, quantification of performance in any given task does not represent a comprehensive evaluation of the computational capabilities of a reservoir.

Assessing the computational capabilities is challenging. In general, different devices will provide different reservoir transformations, with different dynamical regimes of a given device offering further flexibility. To overcome this, task-independent metrics of Kernel Rank (KR), Generalisation Rank (GR) [38], [39], and Linear Memory Capacity (MC) [40] can be employed. These allow empirical measurement of the reservoir's ability to separate, generalise, and remember input respectively. These metrics provide an insight to the properties of a given reservoir configuration along three different computational axes, calculated directly from the transformations the dynamical system provides. The findings of these metrics can be utilised to provide a more informed starting point when optimising these systems to perform machine learning tasks, based on the assumed demands of a given task (e.g., high KR where data is linearly inseparable, high MC for regression tasks with long-term temporal dependencies).

In this paper, we use task-independent metrics to assess the computational capabilities of a modelled interconnected magnetic nanoring array. We show how controlling the scaling and input-rate of data allows these metrics to be tuned, and how their variation correlates to performance in a pair of benchmark classification tasks (spoken and handwritten digit recognition). We then demonstrate how expanding the reservoir's output to include multiple, concurrent measures of the array's magnetic state further improves upon these reservoir metrics and performance in classification tasks.

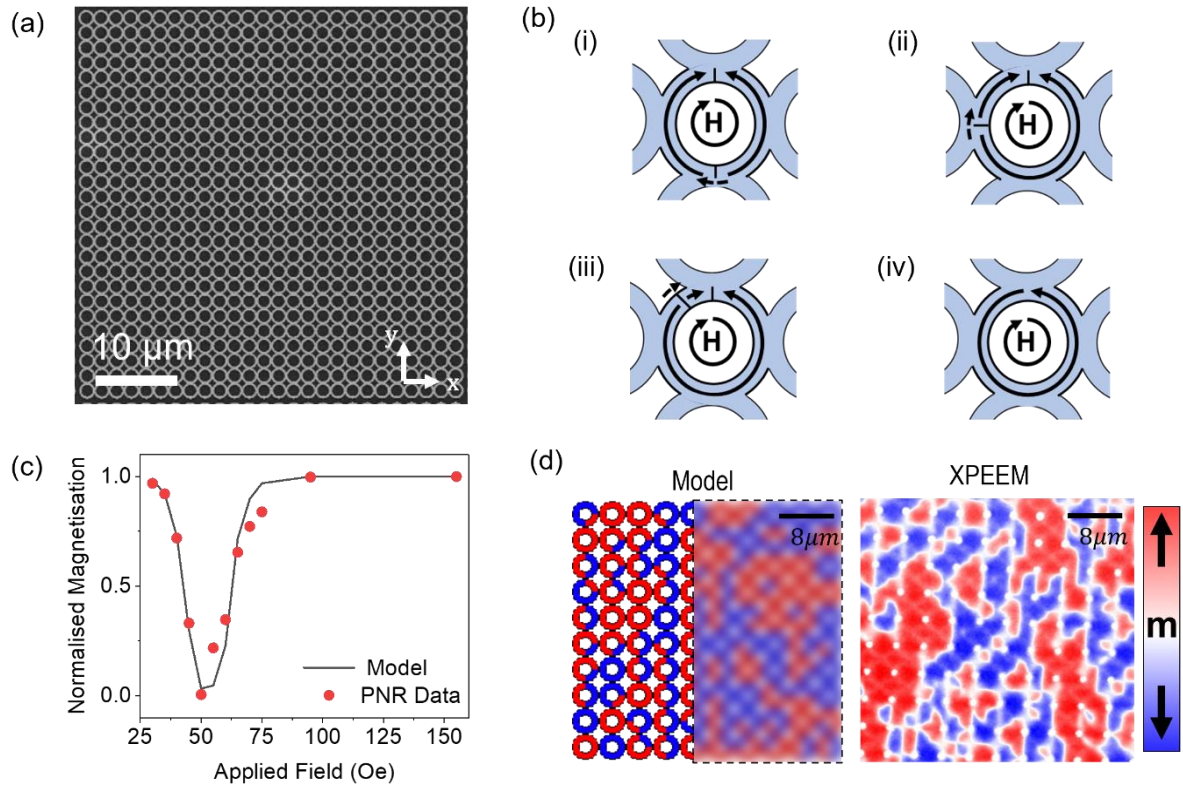

*Figure 2: (a) Scanning Electron Micrograph of a typical nanoring array which has been modelled. (b) Schematics illustrating the DW annihilation process. The upper domain wall stays pinned whilst the lower domain wall moves, leaving the ring in the following states: (i) 'onion' state, (ii) '3/4'state (iii) DW collision, and (iv) formation of a 'vortex' state. (c) Equilibrium*

*magnetisation of the array as a function of applied field as measured by polarised neutron reflectometry (PNR, red symbols) [31] and simulated equilibrium magnetisation given by the fitted model (black line). (d) Magnetisation images generated by the model and by X-ray photoelectron emission microscopy (X-PEEM) after 10 rotations of 50 Oe applied field. Red/Blue colour denotes magnetisation along the axes given by the colour bar. The modelled image includes a blurred region to aid visual comparisons.*

## **Methodology**

*Simulating Magnetic Nanoring Arrays*

The system modelled consisted of a 25 x 25 square array (Figure 2(a)) of 4 μm diameter, 400 nm line width $Ni_{80}Fe_{20}$ rings, with thickness t = 20 nm. This system was experimentally characterised in [31], where we also created and validated a phenomenological model of its behaviour. Here, we used this model, RingSim, to simulate the response of a ring system to streams of data encoded using the amplitude of a rotating magnetic field, with the simulated magnetic states of the array acting as output.

In RingSim, rings existed as either 'onion' states, containing two DWs or 'vortex' states containing no DWs. DWs were instanced into RingSim as agents which attempted to follow the rotating field to minimise Zeeman energy [41]. DWs within RingSim existed as pairs, with one DW instanced as a 'Head-to-Head' DW (H2H, converging magnetisation), and the other as a 'Tail-to-Tail' DW (T2T, diverging magnetisation). When a uniform magnetic field was applied to a ring, the minimum Zeeman energy positions for H2H DWs and T2T DWs were located at opposite sides of the ring, and rotated with the direction of the applied field. For example, for the coordinate system shown in Figure 1(c), magnetisation applied (black arrows) in the positive y direction would have energy minima at the top of each ring for H2H DWs, and the bottom for T2T DWs. Differential movement of DWs as a result of stochastic pinning events (Figure 2(b.i)) led to the formation of '3/4' states (i.e. onion states with one DW displaced by 90°, Figure 2(b.ii)), or collapse of the ring into a vortex state upon DW collision and annihilation (Figure 2(b.iii/iv)).

Junctions between rings created anti-notch-like energy barriers against DW propagation [42], [43]. The size of these energy barriers was modulated by the tangential component of applied field in accordance with Sharrock's equation [33], [44]:

{1} $$\Delta E = E_0 \left(1 - \frac{H_{drive}}{H_0}\right)^\alpha$$

where $\Delta E$ represented the field-modulated energy barrier, $E_0$ was the magnitude of the unmodulated energy barrier, $H_{drive}$ was the component of applied field acting tangentially to the ring at the DW's position, $H_0$ was the zero-temperature depinning field, and $\alpha$ was a geometrical exponent that controls the variation of the energy barrier with applied field. $E_0$ also depended on whether a DW was present in the neighbouring ring on the other side of the junction; where this was the case reducing the energy barrier was reduced by a factor 0.75.

The expected timescale of thermally activated reversal, $t_r$, was calculated from $\Delta E$ barrier using the Arrhenius-Néel law, [45]:

{2} $$t_r = t_0 e^{\frac{\Delta E}{k_B T}}$$

where $t_0$ represented the inverse of the attempt frequency (~ 1 GHz for $Ni_{80}Fe_{20}$ [33]), $k_B$ represented the Boltzmann constant, and T was the temperature.

The rotating magnetic field was modelled as series of discrete steps of π/8 radians. At each step, the field was held for a duration of $t_H = 1/(16*f)$ seconds, where f was the frequency of rotation, taken here to be 5 Hz, of the order of the rotational frequency used in our experiments. For each of these field steps, the probability, P of a DW depinning from by the energy barrier was calculated using:

{3} $$P = 1 - e^{-t_H/t_r}$$

The stochastic nature of DW pinning was modelled by comparing random floating points to pinning probability P. If the generated random floating point exceeded P, then the DW was considered free to propagate via the shortest path to either the appropriate Zeeman energy minima, or an intermediate junction. Interactions between DWs were introduced phenomenologically: If a H2H/T2T DW collided with T2T/H2H DW in the same ring, both DWs were annihilated, leaving the ring in an onion state; If a DW passed a junction where a no DW was present in the neighbouring ring it nucleates a H2H and T2T DW pair into that ring.

RingSim was defined by 4 free parameters: $H_0, E_0, \alpha$, and $\Delta H_0$. The first three parameters were defined in the previous equations, and the final parameter, $\Delta H_0$ represented the standard deviation of a gaussian distribution of $H_0$ across the junctions of the array and was included to approximate the variation in junction properties arising from material or lithographic defects. In this study, these parameters were fitted to the results of polarised neutron reflectivity (PNR) measurements of the array shown in Figure 2(a), where the array's net magnetisation along the y-axis was measured as a function of rotating field amplitude. The PNR measurements were taken following 50 rotations at each applied field amplitude, with saturation and relaxation of the array occurring before each measurement [31]. The data thus represented a dynamic equilibrium magnetisation state of the array at each applied field amplitude.

Figure 2(c) presents PNR data, along with the fit produced by RingSim ($H_0 = 14.25\ mT, E_0 = 1.05eV$, $\alpha = 1.1,\ \Delta H_0 = 1.25mT$). The model fitted the data well, with agreement being particularly strong in the region H$_{drive}$ = 35 - 70 Oe, where the system exhibited an emergent response. Furthermore, images generated from RingSim in this regime of behaviour showed good qualitative agreement with X-ray photo-electron emission microscopy images (X-PEEM), (Figure 2(d)), with both showing extended magnetic domains forming over similar length scales in the array. Further details of the validation of RingSim against experimental results can be found in [31].

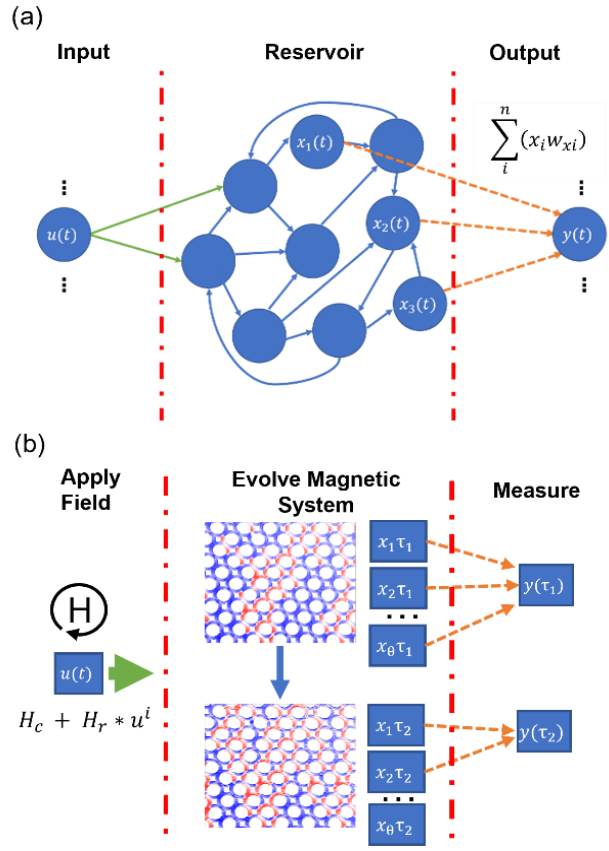

*Figure 3: (a) The ESN approach to RC, showing the layered structure of the model: a fixed reservoir layer is provided with time-varying input via weighted connections. A linear output layer then provides a weighted sum of activities from nodes within the reservoir layer. (b) Our approach, where the ring array acts as a single dynamical node, into which time-multiplexed data is input using a rotating magnetic field, and output is extracted by measuring magnetic properties of the array at the end of each time-multiplexed input.*

### Simulating Reservoir Computing with Ring Arrays

RC involves the transformation of discrete-time input signals, $u(t)$, to reservoir states, $x(t)$ (Figure 3(a)). The reservoir configuration employed here follows the paradigm of a single dynamical node, as introduced by Appeltant et al. [5] (Figure 3(b)). Here, the network was constructed of 'virtual' nodes, created by observing a physical property of a dynamic system as it responds to time-multiplexed input. This approach has been used in a wide range of physical reservoirs due to its ease of implementation [5], [6], [10]–[12].

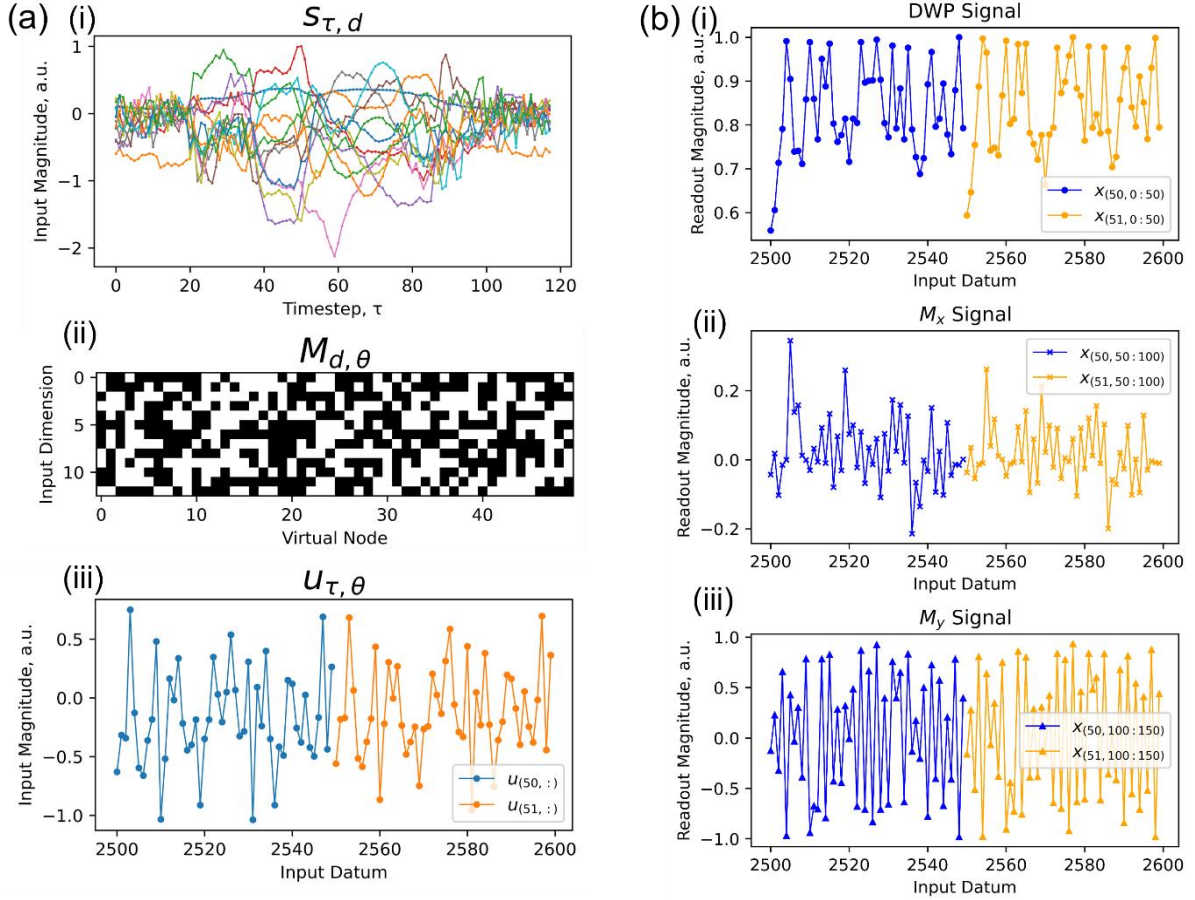

*Figure 4- (a) Outline of the masking procedure for data representing an utterance of the digit 'zero', showing filtered input signal **s** (top, different colours represent different input dimensions d), which is combined with binary input mask **M** (middle), to form masked input signal **u** (bottom, masked signals shown for τ=50/51). (b) Example signals for the three state readout variables, DWP (top), $M_x$ (middle), and $M_y$ (bottom) when driven with $u_{50}$ (blue) and $u_{51}$ (orange) of a spoken digit 'zero'.*

The time-multiplexed input matrix, $u_{\tau,\theta}$, (Figure 4(a.iii)) was generated from raw input signal $s_{\tau,d}$ (Figure 4(a.i)) multiplied with mask matrix $M_{d,\theta}$ (Figure 4(a.ii)), shown the equation below,

{4} $$u_{\tau,\theta} = s_{\tau,d} * M_{d,\theta}$$

where $d$ = number of dimensions of the input signal, $\tau$ = the number of time-steps of the input signal, and $\theta$ = the number of virtual nodes. Matrix $u_{\tau,\theta}$ was then flattened by concatenating row by row, producing a 1D input signal $u_n$ of length $\tau * \theta$. For $d > 1$, $M_{d,\theta}$ was filled with random binary digits to provide different mixtures of the input dimensions across the virtual nodes (figure 4(a.ii)). For single-dimensional inputs, a mask of random floating points was used instead to excite different responses in the reservoir over time.

For an input datum $i$ from $u_n$, the applied field amplitude $H_{applied}^i$ was given by:

{5} $$H_{applied}^i = H_{centre} + H_{range} * u_i$$

where $H_{centre}$ and $H_{range}$ represented the offset and scaling of the rotating field sequence. Each input was applied for a given number of quarter-rotations of field, where $N_q$ denoted the number of quarter-rotations (with quarter-rotations chosen to reflect the fourfold rotational symmetry of the array). Here, values of $N_q$ were chosen such that $N_q$ was smaller than the number of quarter rotations required to reach an equilibrium state. This connected the states of the virtual nodes to one another by maintaining the reservoir in transience [5], [12].

Three variables correlated to the magnetic state of the array were logged at the end of each input, the number of DWs currently in the system, normalised to the number found at saturation (Domain Wall Population, DWP, Figure 4(b.i)); and the array's net magnetisation components in the x and y directions, $M_x$ and $M_y$, Figure 4(b.ii/b.iii)). These were concatenated, producing a reservoir state vector $x_{3n}$ which was three times the length of $u_n$. We note that while these output variables were chosen as they were easily available in RingSim, they are all potentially accessible in electronic measurements of real devices. For example, values representative of $M_x$ and $M_y$ could be measured in spin valve stacks exhibiting GMR multilayers with appropriately aligned pinned layers [20], [36], [37], while a proxy for DWP could be obtained through anisotropic magnetoresistance (AMR) measurements [21], [46], [47].

The output from the reservoir, $Y_{\omega,n}$, was constructed by combining the reservoir state matrix, $X_{\theta,n}$, with output weights, $W^{out}_{\omega,\theta}$, where $\omega$ reflects the number of output nodes, and $n$ reflects the number of input patterns used to construct reservoir state matrix **X**. Output weights were calculated using an ordinary-least-squares method with Tikhonov regularisation, commonly referred to as 'Ridge Regression' and described by the following equation:

{6} $$\boldsymbol{W}^{out} = \boldsymbol{Y}\boldsymbol{X}^T * (\boldsymbol{X}\boldsymbol{X}^T + \gamma^2 \boldsymbol{I})^\dagger$$

where $\gamma^2$ represented the regularisation parameter, $\boldsymbol{I}$ the identity matrix, and † the Moore-Penrose pseudo-inverse operation. Regularisation was performed by selecting the $\gamma^2$ with highest average classification accuracy on the training set, evaluated across multiple shuffles of the training data.

*Task-Independent Metrics*

We estimated the computational properties of the reservoirs using task independent metrics: KR, GR [38], [39], and MC [40]. KR estimated a reservoir's ability to map distinct inputs to different reservoir states, while GR estimated the ability to generalise noisy versions of the same input to similar reservoir states. Generally, higher KR scores mean a better ability to separate data, while lower GR scores reflect a better ability to generalise data. The score of MC approximates how many inputs in the past the reservoir can reconstruct at a given moment.

To measure both KR and GR, N x 1-dimensional input signals of length M were generated from independent and identically distributed (i.i.d.) floating points uniformly distributed between ±1 and applied to the reservoirs. Here, N = 200 and M = 10. For KR, the sequences were unique and uncorrelated from one another; in GR, the sequences were uncorrelated except for the final three inputs, which were identical for each sequence. Before the sequences were applied to the reservoir, they were combined with a fixed mask of random floating points as described earlier for the case where input dimensionality, d = 1, thus producing N input signals of length M x θ. These were then scaled into rotating field sequences using Equation 5. The reservoir was reinitialised to contain onion states uniformly aligned along +y prior each sequence being inputted.

The final reservoir states following each sequence were generated by taking measurements of DWP, $M_x$ and $M_y$ across all θ virtual nodes, thus generating an output matrix $\boldsymbol{O}_{N,3\theta}$. KR and GR were defined as the ranks of these matrices, and thus represented the number of linearly independent responses produced at the reservoir's output when driven with input signals with the characteristics described above. In addition to generating KR and GR for matrices $\boldsymbol{O}_{N,3\theta}$ containing all three state variables, ranks were also generated for $\boldsymbol{O}_{N,\theta}$ matrices for each of DWP, $M_x$ and $M_y$ alone, in order to highlight the effect on computation of evaluating additional reservoir properties for each input datum.

The ranks of **O** were estimated using singular value decomposition, calculated as the number of singular values above an arbitrary small noise threshold, here 0.1. To alleviate biasing higher metric scores to regions of operation where the readout state variables are numerically higher, all output matrices **O** were normalised against the maximum value in **O** prior to singular value decomposition to provide fair rank estimation between field profiles.

To evaluate the ring array's memory capacity, it was driven with an i.i.d. input $u_i$, where i denoted each input datum before masking. Here, i=550, but the outputs from the first 50 inputs were discarded to wash out any initial conditions from the reservoir's response. We then trained $W^{out}{}_{\omega,\theta}$ to recover past inputs $u_{i-k}$ for each delay k as output $y_k$, with a 250:250 train/ test split. MC was evaluated from the covariance between the delayed inputs $u_{i-k}$ and the trained reconstruction of input $y_k$ for summed across all nodes $\theta$ for each delay via the following formula:

$$\{7\} \qquad MC = \sum_{k=1}^{\theta} \frac{cov^2(u_{i-k}, y_k)}{\sigma^2(u_i)\sigma^2(y_k)}$$

MC provided a basic insight into the memory properties of the reservoir, with MC approximating the number of time-steps in the past over which the network could reliably recall previous inputs.

Heatmaps were generated showing metric values (KR, GR, MC) for a range of driving field parameters $H_{centre}$, $H_{range}$, and $N_q$. In these heatmaps, and the subsequent performance heatmaps for the digit recognition tasks, the scaling parameters were instead expressed as the mean field applied, as well as the standard deviation of the field sequence for a given set of input parameters. This is done to account for the reduction in effective range caused by the masking process, as well as the non-uniform distribution of the task data, allowing more effective correlation between the metrics and the task performance.

*Digit Recognition Tasks*

A pair of benchmark classification tasks were chosen to assess reservoir performance: spoken digit recognition (NIST TI-46 database, [48]) and handwritten digit recognition (MNIST, [49]). Both consisted of a total of 500 utterances/images of the digits 0 – 9. For TI-46, each of five female speakers provided ten utterances of each digit. Inputs were created using a Mel-Frequency Cepstral filter [50] to produce responses in 13 frequency bands across 50 ms windows, generating a raw input signal s with d=13 and τ equal to the number of windows generated by a given utterance. For MNIST, 50 images of each digit were taken randomly from the 'training' set of the database. The [28x28] pixel images were considered as signals where d=τ=28, and multiplexing was performed column-by-column. For both tasks, the array was initialised to contain uniform 'onion' states aligned along +y prior to each input sequence.

Performance was assessed across multiple 80:20 splits of training and testing data for each digit. The 500 datapoints were split into train/test groups randomly, and performance was averaged over 100 different shuffles of the data. For training/testing, the output for every τ in each signal was labelled according to the digit they represent using one-hot encoding, generating state matrix $X_{\theta,m}$ and target matrix $Y_{\omega,m}$ where m was the total number of time-steps across all utterances.

To classify unseen data, the activation for each of the ω=10 outputs was calculated as the cumulative sum across all time-steps τ, meaning that the classification algorithm could handle inputs of different number of timesteps. Classification was performed using a winner-takes-all approach based on the output with the highest activation. We assessed performance for output vectors consisting of each of the three state variables independently, as well as for all three simultaneously

**Results and Discussion**

*Task Independent Metrics*

Figure 5 presents heatmaps for KR, GR and MC created by of varying $H_{centre}$ and $H_{range}$ values for each of the three state variable measurements. Data are presented for a fixed input rate of $N_q$ = 2, which exhibited best overall performance in the tasks, while heatmaps for input rates $N_q$ = 1 and $N_q$ = 4 can be found in the supplementary material (Figures S1 and S2). Each metric was bounded by number of nodes, here 50, as the maximum rank of output matrices is limited to the smallest dimension of $\mathbf{O}_{N,\theta}$, and MC is similarly bounded as the maximum covariance for a given delay is 1, summed across all nodes $\theta$. The three input rates show broadly the same behaviour, with some shifting peak values towards higher mean fields and ranges for the fastest input rate of $N_q$ = 1 since higher fields perturbed the system more significantly when each input was applied for less time.

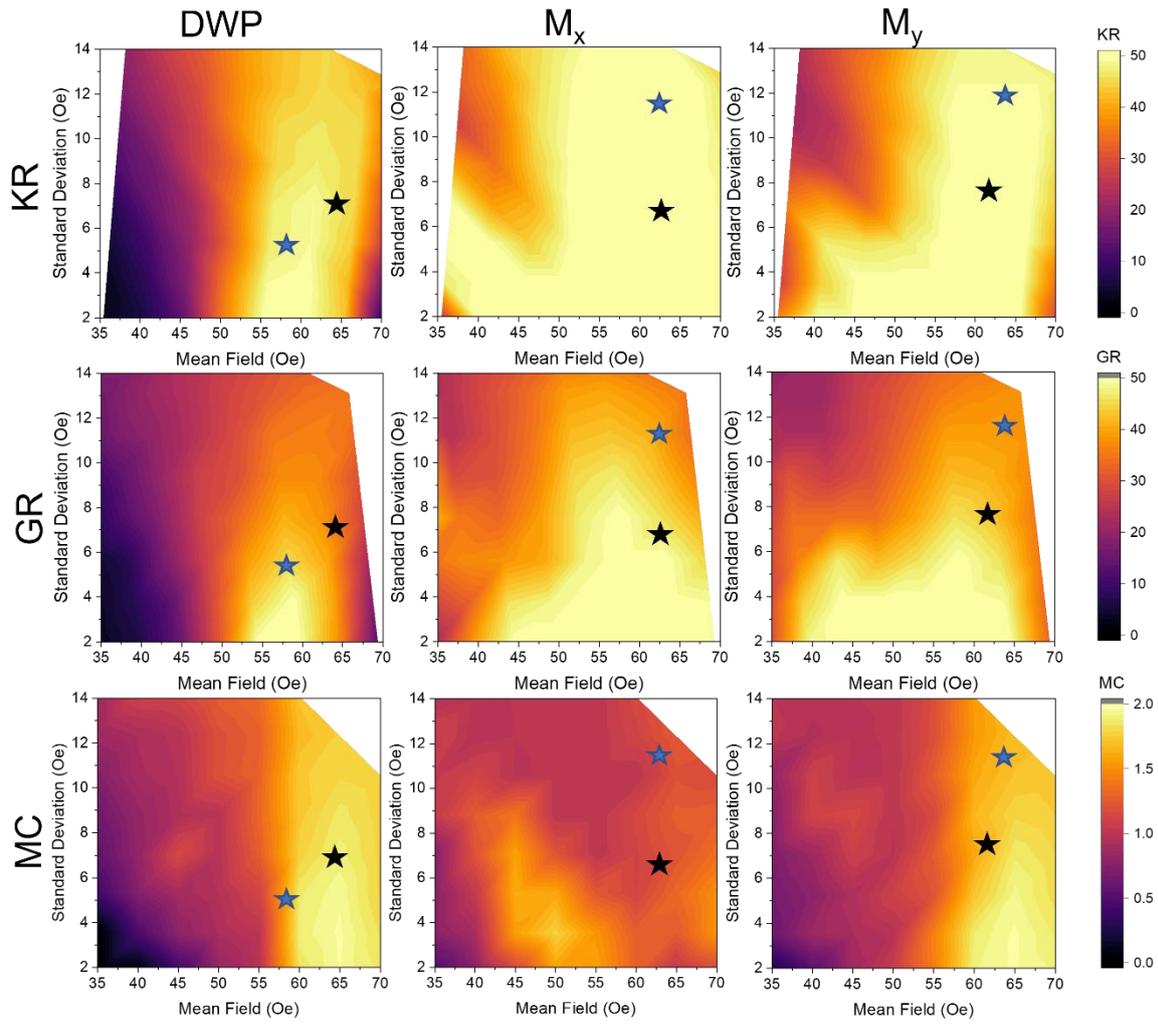

*Figure 5- Empirically-measured reservoir metrics Kernel Rank (KR), Generalisation Rank (GR), and Memory Capacity (MC) for an input rate of $N_q$ = 2, or a half rotation of field per input, for each of the output parameters of Domain Wall Population, X Magnetisation, and Y Magnetisation taken independently. Black/blue stars represent peak performance in spoken/handwritten digit recognition for each reservoir state variable, while metric scores are reflected according to the colour bars on the right.*

A wide distribution of different reservoir properties were observed in the heatmaps. The configurations most suitable for reservoir computing were clustered around mean field = 50-70 Oe, with KR reaching its maximum value and MC > 1, indicating the presence of both nonlinearity and memory in the reservoirs' responses.

Useful reservoirs were not expected in the 'hotspots' of relatively high GR, as in those regions the reservoirs could not map similar input sequences to similar output states. These regions likely corresponded to dynamical regimes where the inherent stochastic noise of the system was large and obscured the underlying signal, preventing effective generalisation. For all three state variables, there was a wide region of high KR. This represented regimes where the reservoirs' dynamics were suitably non-linear due to a combination of the ring array responding non-linearly to rotating field amplitude (Figure 2(c)) and the connections between virtual nodes produced by the system being maintained transience. Effective reservoir computers were likely to be found in regions where KR > GR as these reflected the system operating in a complex yet ordered state. When optimising task performance, this criterion allows a reduction of the search space without discarding useful reservoir configurations.

All reservoir configurations exhibited low MC, with a maximum MC = 2. This feature can be explained by the time-multiplex approach to RC used: consecutive timesteps of the input data were separated by the entire sequence of virtual nodes and so the memory of the system primarily acted to connect virtual nodes together, rather than connect data from different timesteps. In other time-multiplexed networks with a single dynamical node, a delay line is often included to provide feedback of output from previous timesteps [5], [10], [51], and would likely augment the memory characteristics of the ring array reservoirs similarly. However, this is beyond the scope of this paper, as the goal was to characterise the computational properties of the dynamical system itself, without aid of peripheral feedback methods. As memory is critical to reservoir computation, a decision criterion for restricting the hyperparameter space search as part of task optimisation can be drawn where MC > 1.25 in order to eliminate reservoirs without effective memory.

There were noticeable differences between the metric maps for DWP and $M_x$/$M_y$, with the directional magnetisation components having higher KR/GR values in general. One possible reason for this discrepancy related to the relative complexity of each state variable, since the DWP measure was indifferent to the direction and size of domains in rings, while the magnetisation components were sensitive to these additional factors, and hence exhibited a richer dependence on the system's state. Additionally peak values of MC and KR were maintained at higher input standard deviations for $M_x$/$M_y$. This arose from differences between the expected equilibrium values of DWP and magnetisation for a given field; DWP saturated to maximum values at lower applied fields than magnetisation as the rings form '3/4' configurations (thus maximising DWP) at lower fields than they formed uniformly aligned onion states (thus maximising magnetisation).

There were also differences between the metric maps for $M_x$ and $M_y$ which can be seen most clearly for $N_q$ = 2 (Figure 5) and $N_q$ = 4 (Figure S2) where the regions with high KR and GR extended over larger proportions of the maps for $M_x$ than $M_y$. We suspect these differences occurred because readouts were always performed when the field was aligned along the y-axis for even values of $N_q$, meaning the system was less significantly perturbed in x. Smaller perturbations meant a smaller signal-to-noise ratio for the state variable compared to the inherent noise of the system, and hence higher values of KR and GR, especially when the mean field and standard deviation were lower. $M_x$ also produced a lower peak value in MC than $M_y$. Similarly, this is likely due to $M_x$ having small magnitude, and hence hindering reconstruction with worse noise properties, when operating in the regime where the system seems to show the biggest dependence on past input (Mean Field > 60 Oe, evidenced by peak MC scores for other two variables).

The differing metric distributions for each of the state variables suggested that they provided different data transformations, and thus could be combined to create an enhanced reservoir. This was demonstrated through production of reservoir metric heatmaps for output vectors containing all three state variables together, as shown in Figure 6.

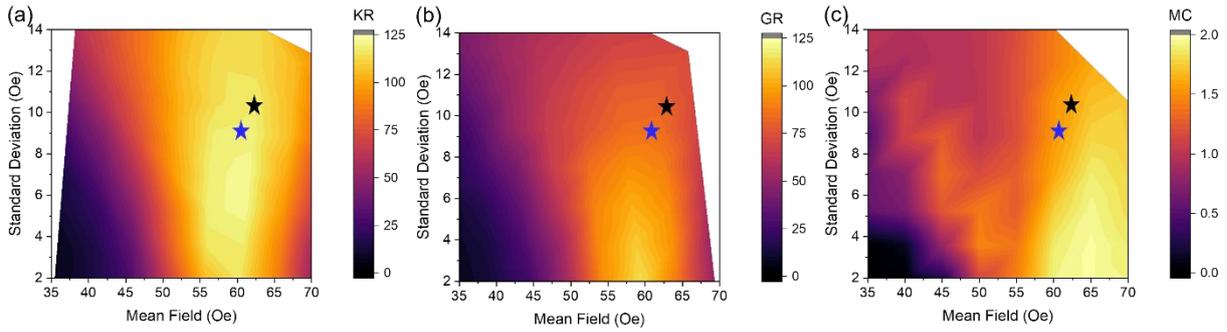

*Figure 6- Three-output-variable metric maps for (a) KR, (b) GR and (c) MC for an input rate of $N_q = 2$. Black/Blue stars represent the highest performing configurations in the spoken/handwritten digit recognition tasks respectively.*

The three-output-variable metric maps showed substantial improvements over those for the individual variables. Both KR and GR increased, highlighting the additional nonlinear mappings provided by using the three state measurements concurrently. However, not all the additional nodes contributed additional nonlinearity. As noted previously, the upper bound for KR and GR was equal to the number of nodes/weights in the virtual network, i.e. 150 when using all three output variables. Both KR and GR peaked below this maximum value, with maximum ranks of 121 and 117 respectively, illustrating the diminishing returns of adding additional virtual nodes. The maximum MC of the system did not increase. This was likely due to the three state variables having similar rates of change with respect to changing inputs.

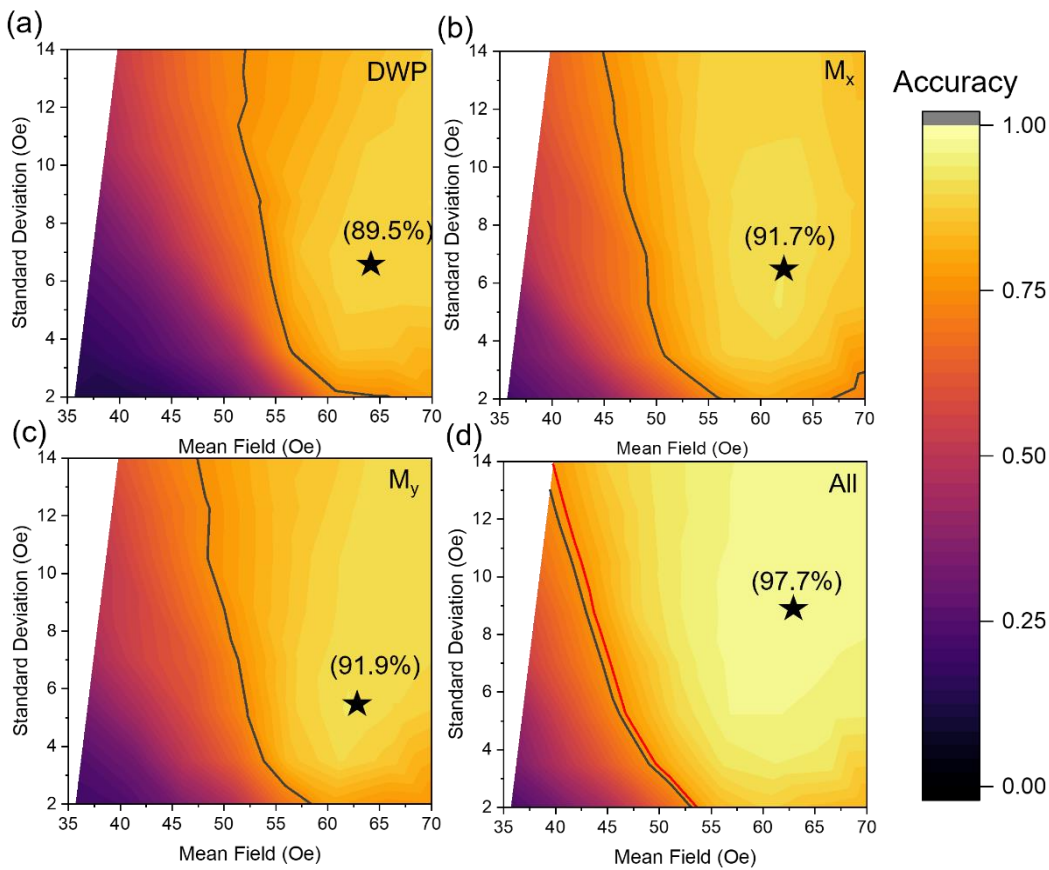

*Figure 7- Test accuracy for TI-46, 100 utterances by five different speakers, 80:20 training: testing split. Quoted accuracies are for 100 different shuffles of training/test data. Four maps represent which outputs constituted the features used for classification: (a) DWP output (b) $M_x$ (c) $M_y$ (d) combined outputs. Blacks line = performance without reservoir transformation, where $\theta = 50$. Red line = baseline performance for input data generated from a mask with $\theta = 150$.*

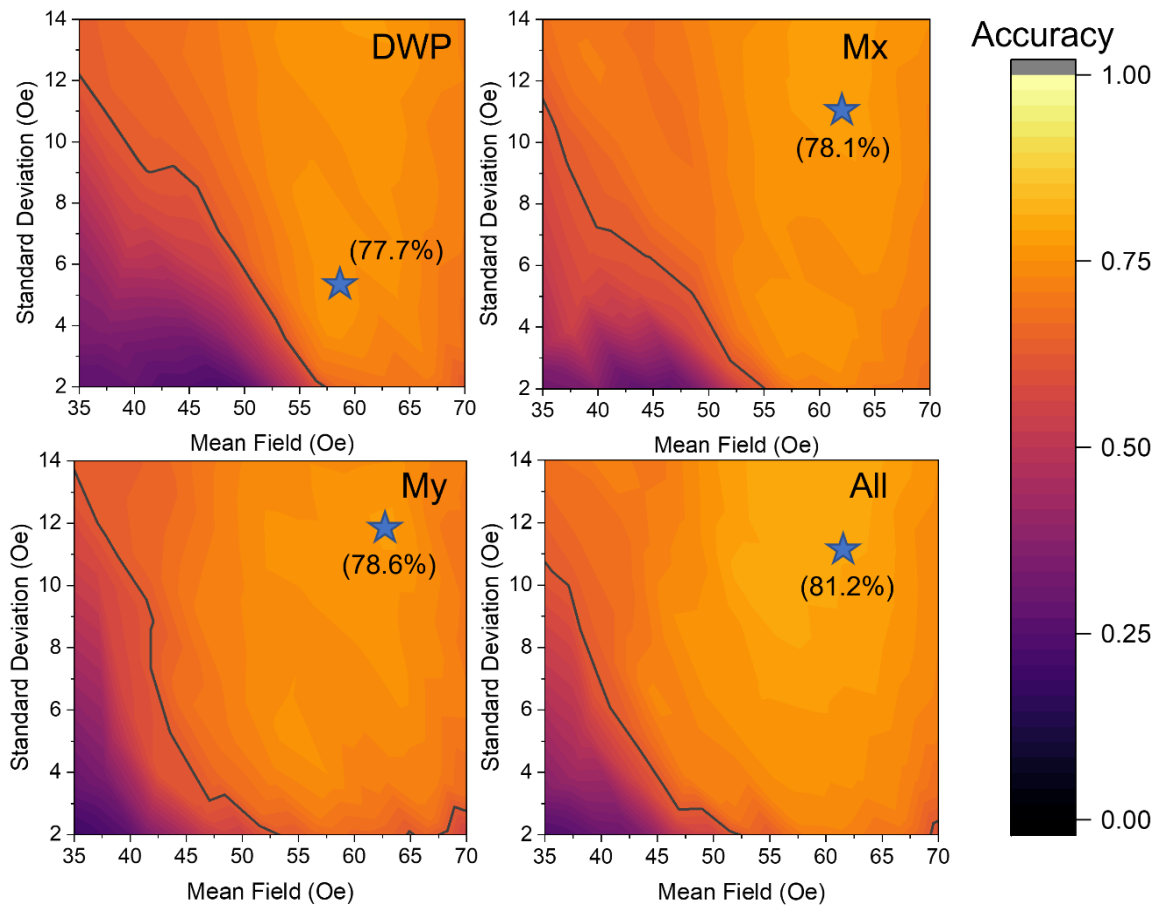

*Figure 8- Task performance for 100 different shuffles of training/testing data of the MNIST task. Blue stars and percentages represent peak classification accuracies for each configuration. Black line represents control performance generated by skipping reservoir transformation, where θ = 50.*

*Digit Recognition Tasks*

Figure 7 presents maps of TI-46 performance for each of the output variables, as well for all three combined. Peak accuracies were 89.5%, 91.7%, and 91.9% for the DWP, $M_x$ and $M_y$ outputs respectively and increased to 97.7% for the three properties combined. As a control, performance tests were performed where the output weights were trained directly on the masked input data, thus eliminating the reservoir's transformation. Control measurements were created for the three-output-variable case by generating masked inputs with $\theta$ = 150 to ensure parity in trainable parameters. The 50 and 150 virtual node control setups achieved average accuracies of 75.8% and 77.2% respectively. Thus, all the output configurations substantially outperformed the control. Reservoir configurations which outperformed the control data are bounded on the heatmaps by black and red lines, showing 50 and 150 node controls respectively. Peak accuracies were competitive with proposed architectures with similar pre-processing (94.8% with 500 training samples and 50 virtual nodes on a superparamagnetic array [10], 99.8% with 900 training samples and 400 virtual nodes on a spin-torque nano-oscillator [52]).

The sequential MNIST handwritten digit recognition task was performed similarly, with both individual state variables, as well as the three outputs combined (Figure 8). Again, each state variable provided considerable improvement over the control configuration, with scores of 77.7%, 78.1%, and 78.6% for the DWP, $M_x$ and $M_y$ outputs respectively, compared to a control accuracy of 61.2%. Performance in this task rivalled an ESN with slightly fewer nodes (100 nodes, 79.43% accuracy [53]). In both tasks,

$M_y$ slightly outperformed the other two output measures, owing to its greater expressivity than DWP, and improved signal-to-noise properties compared to $M_x$, properties which are reflected in the metric maps where $M_y$ exhibited a higher KR than DWP, and a lower GR than $M_x$.

While both tasks showed an improvement in performance when the concurrent state variables were combined, the MNIST task showed a smaller increase in performance compared to the TI-46 task, rising to a peak accuracy of 81.2%. The task independent metrics can explain the discrepancy between the improvement provided to the two tasks; metric evaluation showed a gain in the KR and the GR of the system with combined outputs, but no change to the system's MC. The sequential MNIST task requires effective correlations to be drawn over long separations for successful classification (e.g., the left-most, 'earlier' columns for the digit '3' are crucial to avoid confusion with an '8'), which require longer memory capacities than were provided by the ring array. Adding the additional readouts did not improve memory capacity, hence there was a limited improvement to performance.

Strong correlations were observed between the reservoir metric maps and task performance maps. The regions of highest performance in both digit recognition task (indicated by symbols on the metric plots in Figures 5 and 6) were found at points that had high KR scores but avoided areas with high GR. MC was also strongly correlated to performance, with the highest performing reservoirs all having a memory capacity above 1.5. This indicated the importance of having both memory and nonlinearity in the system to provide useful transformations for classification. Practically, a hyperparameter search to optimise these reservoirs for a task could be confined within the decision boundaries outlined in the task-independent metrics section: KR – GR > 10, and MC > 1.25. This would reduce the parameter space of the original search considerably whilst still capturing the peak performance in both tasks. Decision boundaries overlayed on task performance heatmaps are shown in Figure 9.

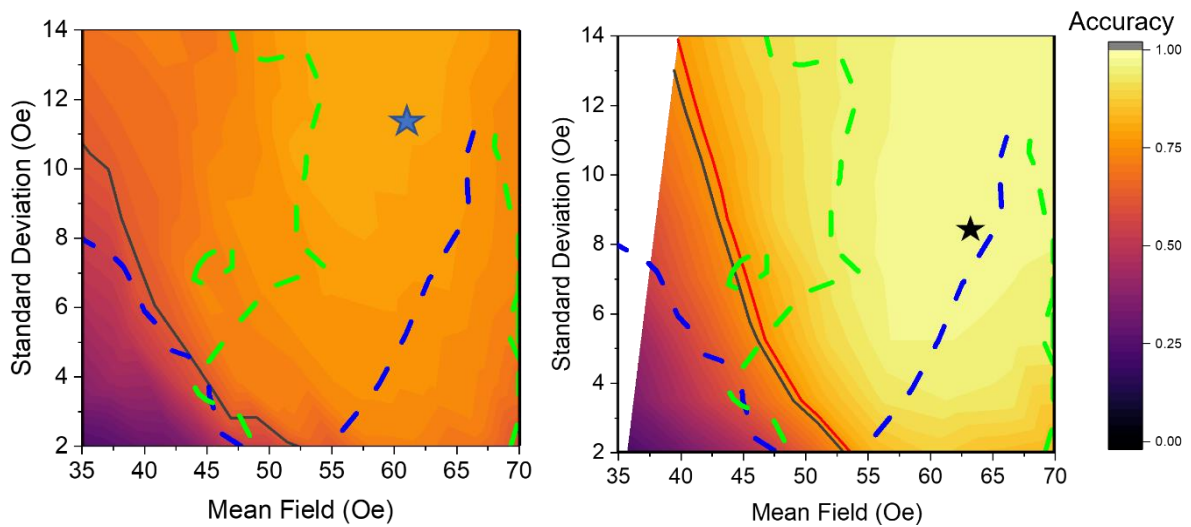

*Figure 9- Decision boundaries of KR – GR > 10 (blue) and MC > 1.25 (green) overlayed on performance heatmaps for the MNIST task (left) and TI-46 task (right), showing peak performance is captured for both tasks by these decision boundaries.*

In conclusion, we have shown that reservoirs based on interconnected magnetic nano-rings arrays can achieve a broad range of signal transformations, and explored the suitability of these for reservoir computing by calculating task independent metrics KR, GR, and MC. We then showed how the range of available metrics could be expanded by taking multiple concurrent measurements of the system's magnetic state. Finally, we demonstrated that these metrics correlated to performance in classification tasks and highlighted the substantial increase in performance that using the additional measurements of system state brought. The memory capacity of the ring ensembles was limited, primarily due to the time-multiplexed approach to MC we adopted here. However, as the magnetisation dynamics of the ring ensembles are spatially distributed, they are naturally well-suited to spatially multiplexed approaches where local data inputs are used to address discrete regions of the array as physical, rather than virtual, nodes. This should allow substantial enhancement of the arrays' memory characteristics. Our work represents an important step towards realising RC in magnetic ensembles with emergent magnetisation dynamics.

**Acknowledgements**

The authors thank STFC for beam time on the Offspec beamline at the ISIS Neutron and Muon Source (https://doi.org/10.5286/ISIS.E.RB1810656) and on beamline I06 at the Diamond Light Source, and thank Jordi Prat, Michael Foerster and Lucia Aballe from ALBA for providing quadrupole sample holders[54]. R.W.D., T.J.B., and I.T.V. acknowledge DTA-funded PhD studentships from EPSRC. The authors gratefully acknowledge the support of EPSRC through grants EP/S009647/1, EP/V006339/1 and EP/V006029/1. This project has received funding from the European Union's Horizon 2020 FET-Open program under grant agreement No 861618 (SpinEngine).

**Data Availability**

The data that support the findings of this study are openly available in ORDA, at https://figshare.shef.ac.uk, reference number 10.15131/shef.data.16904806.